\documentclass[aps,longbibliography,twocolumn,nofootinbib]{revtex4-1}
\usepackage[colorlinks,citecolor=blue,linkcolor=blue,urlcolor=blue]{hyperref}
\usepackage[mathscr]{eucal}
\usepackage[latin1]{inputenc}
\usepackage{amsmath}
\usepackage{amsfonts}
\usepackage{amssymb}
\usepackage[usenames,dvipsnames]{color}
\usepackage[pdftex]{graphicx}
\usepackage{graphicx}
\usepackage{hyperref}

\newcommand{\C}{\mathbb{C}}

\usepackage{enumerate}

\usepackage{colordvi}
\newcounter{mnotecount}[section]

\newcommand{\comment}[1]{}
\newcommand{\be}{\nopagebreak[3]\begin{equation}}
\newcommand{\ee}{\end{equation}}
\newcommand{\ba}{\nopagebreak[3]\begin{eqnarray}}
\newcommand{\ea}{\end{eqnarray}}

\begin{document}
 \title{Lorentzian Connes Distance, Spectral Graph Distance and Loop Gravity}
\date{\today}

    \author{Carlo Rovelli}
\affiliation{CPT, CNRS UMR7332, Aix-Marseille Universit\'e and Universit\'e de Toulon, F-13288 Marseille, EU}

%%%%%%%%%%%%%%%%%%%%%%%%
%%%%%%%%%%%%%%%%%%%%%%%%

\begin{abstract}                
% \vskip2mm
\noindent 
Connes' formula defines a distance in loop quantum gravity, via the spinfoam Dirac operator. A simple notion of spectral distance on a graph can be extended do the discrete Lorentzian context, providing a physically natural example of Lorentzian spectral geometry, with a neat space of Dirac operators. The Hilbert structure of the fermion space is Lorentz covariant rather than invariant. 

%\vspace{15mm}
%\vspace{20mm}
\end{abstract}

%%%%%%%%%%%%%%%%%%%%%%%%
%%%%%%%%%%%%%%%%%%%%%%%%
\maketitle
%%%%%%%%%%%%%%%%%%%%%%%%
%%%%%%%%%%%%%%%%%%%%%%%%

\section{Connes distance}

Bernhard Riemann defined geometry in term of a metric tensor $g_{ab}$ on a manifold, with the distance between two points $p$, and $q$ given by
\be
d(p,q)={\rm Inf}_\gamma\int_\gamma \ \sqrt{g_{ab}dx^adx^b},
\ee
the minimisation being over all lines $\gamma$ that start at $p$ and end at $q$.   Alain Connes has studied an equivalent definition of distance: 
\be
d(p,q)={\rm Sup}_f|f(p)-f(q)|,   \ \ \  |[D,f]|\le1,
 \label{distance}
\ee
the Sup being over all functions $f$ on the manifold whose commutator with a derivative operator $D$ has (sup) norm less than unity.  The metric structure of the manifold is coded in $D$ rather than $g_{ab}$ \cite{Noncommutative}.  

The logic of Connes's definition is simple: the distance between two points is equal to the maximal variation of a function between the two points, if the derivative of the function is less than one (Figure 1). 
\begin{figure}[b]
\includegraphics[height=3.5cm]{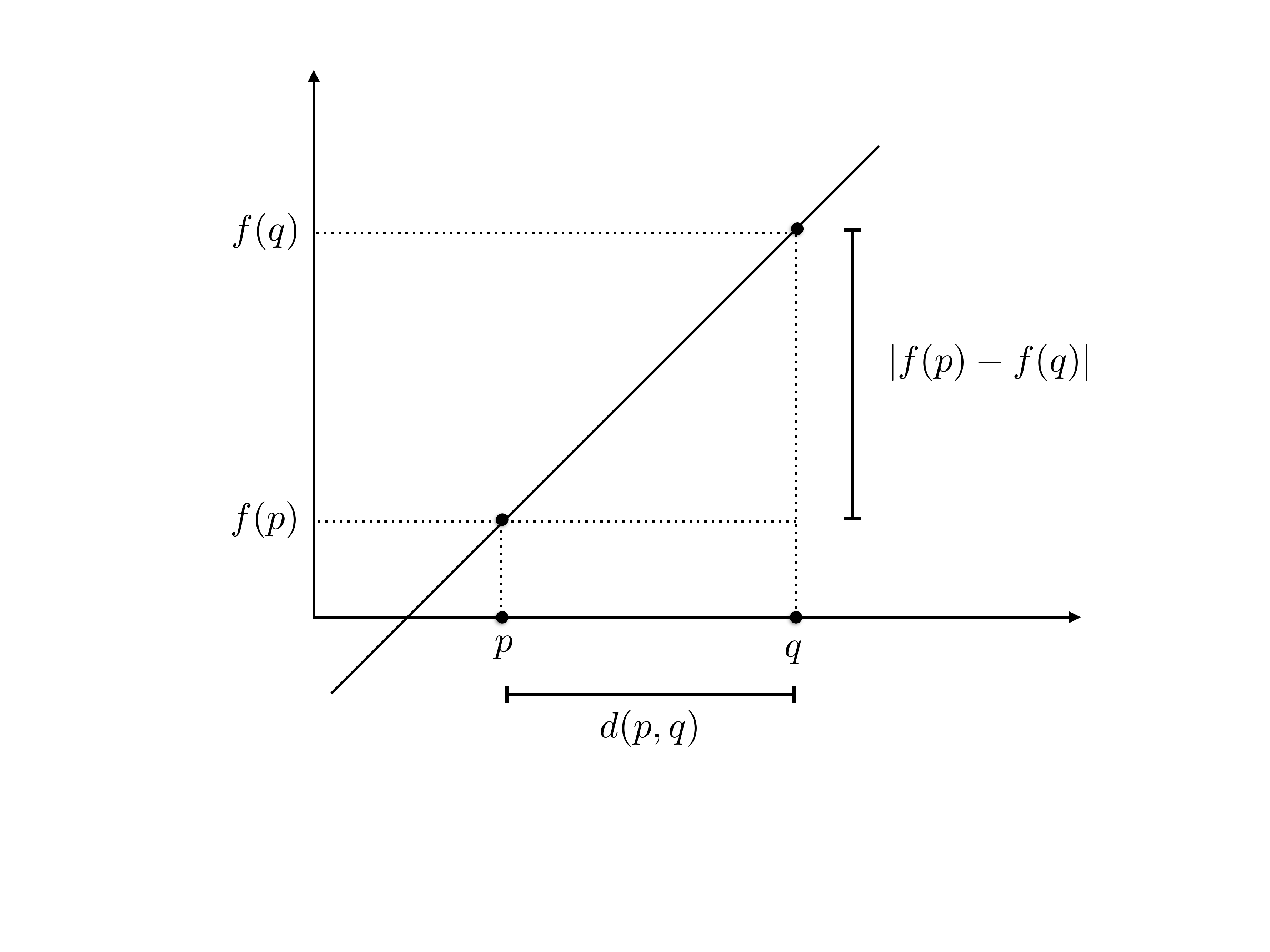}
\caption{Connes distance: if the derivative of $f$ is one, then $d(p,q)=|f(p)-f(q)|$.}
\end{figure}

Connes has shown that the Dirac operator $D$ acting on the fermion fields on a metric (spin) manifold captures --in an elegant and compact manner-- the geometry of the manifold.   He has proven that a compact Riemannian space can be reconstructed from the algebraic structure formed by the Dirac operator acting on the Hilbert space $\cal H$ of the Dirac fields and the commutative algebra $\cal A$ of the functions $f$ on the manifold (acting multiplicatively on the fermion fields). The triple $C=({\cal A},{\cal H},D)$ is called a ``spectral triple".  As pointed out by Gelfand, the points of the manifold can  reconstructed from the algebra as the maximal ideals ($p(f)=f(p)$); the metric structure is determined by \eqref{distance}.  Connes' notion of distance is powerful because it generalises to  wider contexts than Riemann's, including non--commutative geometry.  

In this note, I observe that Connes construction can be used to define distance in the context of a graph, and, more interestingly, in loop quantum gravity (see \cite{Rovelli:2004fk,ThiemannBook,Ashtekar:2013hs,Han:2013tap} and \cite{Rovelli} for a recent introduction), using the Dirac operator defined by the fermionic loop action  constructed recently in \cite{Bianchi:2010bn,Han:2011as}.  

Distance (as opposite to area and volume) has so far been puzzling in loop gravity \cite{ThiemannLength,Bianchi:2008es}.  Connes construction provides a novel way of talking about four-dimensional distance in the covariant version of the theory.\footnote{On the relation between LQG and spectral geometry, see Marcolli and van Suijlekom \cite{Marcolli2013}. An interesting attempt to associate (Lorentzian) spectral triples to spin-foams, and relate the Dirac operator with the the Tomita-Takesaki modular generators has been made by Bertozzini  \cite{bertozzini2010}. The relation with the construction here deserves to be explored. See also \cite{Aastrup2012}.}

The loop-gravity realisation of a spectral geometry is Lorentzian and forms a  generalisation of the Euclidean definition of spectral triple. As we shall see, $D$ and $\cal A$ act naturally on a space $\cal H$ which has a Lorentz covariant \emph{family} of Hilbert structures. This appears to be the natural physical structure.\footnote{The Hilbert space is replaced by a Kre\u{i}n space. Generalisations of the spectral theory of this kind in order to accomodate Lorentzian geometry have been considered in particular by Strohmaier  \cite{Strohmaier2001}, see also \cite{Paschke2006,Dungen2009,Borris2009,Verch2011,Franco2013,Franco2014}.  The generalisation of Connes distance to the Lorentzian domain is  investigated in \cite{Parfionov2000,Moretti2003}.}

The discrete Dirac operator turns out to be determined by the assignment of a covariant Minkowski vector $n_e$ to each edge $e$ of a graph.  Geometrically, $n_e$ is the 3d volume three-form associated to a three-cell of a triangulation dual to the graph.  This characterization of a discrete Lorentzian geometry is related to that recently explored in  \cite{Wieland2014} and \cite{Cortes2014}, where the vector $n_e$ plays a central role as well.   The construction provides a neat definition of a space of Dirac operators.

\section{Graph spectral triple}

As a preliminary exercise, consider the minimal version of commutative spectral geometry defined by a graph (see also \cite{Iochum1999}).  A natural notion of distance between two nodes on a graph is given by the minimal number of links forming a path connecting the two nodes.  A strictly related notion of distance is provided by a finite Connes' spectral triple, as follows. 

Consider a finite abstract graph $\Gamma$. This is defined by a finite set $\cal V$ with $V$ elements $v,w,...$ which we call vertices, and a set $\cal E$ of $E$ ordered pairs $e=(v,w)$ which we call edges.  Assume for simplicity that there is at most one edge per each pair of vertices. Such a graph is uniquely defined by its adjacency matrix: an antisymmetric $V\times V$ matrix with matrix elements $\Gamma_{vw}$ defined by 
\be
\Gamma_{vw}=-\Gamma_{wv}=1,\ \ \ \  {\rm if}\ (v,w)\in{\cal E}
\ee
and otherwise vanishing.  Associate a space ${\cal H}_v=\C$ to each vertex $v$, and consider the space ${\cal H}=\oplus_v {\cal H}_v$; the natural basis in this space can be written as $|v\rangle$.  Let then $\cal A$ be formed by all complex functions $f: {\cal V}\to \C$. Finally, consider the operator $D$ acting on ${\cal H}$ defined  by
\be
\langle v|D|w\rangle=i\Gamma_{vw}.
\ee
Given a norm, the Connes's distance formula \eqref{distance} defines the distance between any two nodes of the graph. The simplest possibility is to take the norm $|A|_\mu=max_{v,w}|A_{vw}|$, which, as we shall see below, works excellently. A more invariant choice is the maximal eigenvalue operator norm.

Let us see how this works in the simplest case.  Consider the distance between two vertices $v_1,v_2$. According to \eqref{distance}, the distance between the two is given by the Sup of $|f(v_1)-f(v_2)|$ among all functions $f(v)$ such that the norm of  $[D,f]$ is less than one. The non vanishing matrix elements of the commutator are 
\be
[D,f]_{vw}=i\Gamma_{vw}f_w-f_v i \Gamma_{vw}= i(f_v-f_w).
\ee
for all $(v,w)$ forming a link.  If we take the $|\ |_\mu$  norm we have immediately the result that the function cannot change by more than one across any link. The distance if therefore given immediately by the minimal number of links between the vertices.\footnote{If we take the max-eigenvalue norm, an eigenstate of the commutator is $|\psi\rangle=|v_1\rangle +i|v_2\rangle$. Its eigenvalue is $f_{v_1}-f_{v_2}$. Therefore we can satisfy and saturate the inequality $ |[D,f]|\le1$, posing
\be 
f_{v_2}-f_{v_1}=1.
\ee
Therefore the distance between two linked points is again one. Up to an overall factor $\sqrt2$, this extends to more distant points. The reader can simply work out other examples, illuminating the relation with the continuous commutative spectral triples.}

\section{Distance in covariant loop quantum gravity}

In the covariant formulation of loop quantum gravity, fermions are described as follows. Let ${\cal C}$ be a two-complex. Its vertices $v$ and edges $e$ form a graph.  A chiral fermion field is represented by a Weyl spinor $\psi_v\in \C^2$ associated to each vertex $v$.  Therefore a linear space ${\cal H}_v=\C^2$ is associated to each vertex.  (The generalisation to a Dirac fermion is immediate.)  Let then ${\cal H}=\oplus_v {\cal H}_v$.

It is important to observe that the space ${\cal H}_v=\C^2$ carries the fundamental representation of $SL(2,\C)$, the double cover of the Lorentz group, and this representation is not unitary. 
It carries also the fundamental, spin-$\frac12$, \emph{unitary} representation of $SU(2)$. The scalar product that makes $\C^2$ into the \emph{Hilbert} space of this unitary representation is 
\be
\langle \psi |\phi  \rangle =\bar\psi^1\phi^1+\bar\psi^2\phi^2.
\label{sp}
\ee
and is  \emph{not} invariant under $SL(2,C)$. The transformation properties of this scalar product under $SL(2,\C)$ can be neatly expressed,  
using the four Pauli matrices $\sigma^I\!=\!(1\!\!1,\vec\sigma),\ I=0,1,2,3$,  by writing \eqref{sp} in the form 
\be
\langle \psi |\phi  \rangle = \langle \psi |\sigma^0|\phi  \rangle;
\ee 
then scalar products in the family 
\be
\langle \psi |\phi  \rangle_n :=\langle \psi |n_I\sigma^I|\phi  \rangle
\ee
 are each invariant under the $SU(2)$ subgroup of $SL(2,\C)$ which (in the vector representation) leaves the covariant vector $n=\{n_I\}$ invariant. Under a change of Lorentz frame, they transform into one another. 

Now, in the partition function of loop quantum gravity, the geometry associated to the two-complex is determined by an $SL(2,\C)$ group element $g_{ve}$ associated to each couple vertex-edge and a ``volume" $v_e\in R^+$ associated to each edge. These data determine a covariant Minkowski vector $n_e$ associated to each edge $e$, via 
\be
n_e=v_e\ g_{ve}\sigma^0 g^{-1}_{v'e} =(n_e)_I\sigma^I .
\ee 
where $v$ and $v'$ are the source and target of $e$.  The geometry is  captured by a covariant Minkowski vector $n_e$, or, equivalently, a linear combination of Pauli matrices $n_e=(n_e)_I\sigma^I$, to each (oriented) edge $e$ of the two-complex.  
 The Dirac action appears in the following form in the amplitude of the theory
\begin{eqnarray}
S&=&i\sum_e\ v_e\  \langle \psi | g_{ve}\sigma^0 g^{-1}_{v'e}| \psi \rangle\nonumber\\
&=&i\sum_e\   \langle \psi |n_e| \psi \rangle.
\label{dirac}
\end{eqnarray}
The spinfoam Dirac operator is therefore 
\be
\langle \psi |D| \phi \rangle =  \ i\!\!\!\sum_{e=(v,v')}  \langle \psi_v |n_e| \phi_{v'} \rangle
\label{D}
\ee
which is a dressed version of the graph Dirac operator considered in the previous section. The dressing, given by the $n_e$, yields the Lorentzian structure.  

Then, \eqref{distance} where $f$ is a functions $f(v)$ on the vertices, provides a definition of distance on the two complex.   This is the definition of distance in the spinfoam context that we were seeking. 

The formal relation between $n_e$ and the gravitational field is the following. Choose a cellular decomposition of spacetime, associate a vertex to each cell and an edge $e$ to each couple of adjacent cells. Then 
\be
    (n_e)_I=\epsilon_{IJKL}\int_{\Sigma_e} e^J\wedge e^K\wedge e^L
\ee
where $e^I=e^I_adx^a$ is the tetrad field (related to the metric by $g_{ab}(x)=e^I_a(x)e^J_b(x)\eta_{IJ}$, where $\eta$ is the Minkowski metric) and $\Sigma_e$ is the 3d surface separating the two cells joined by $e$.  Compare this with the continuous Dirac action
\ba
    S&=&\int d^4x\ \det{e} \ \bar\psi \gamma^I e^a_I D_a \psi 
   \nonumber \\  &=& \int 
      \epsilon_{IJKL}\   \bar\psi \gamma^I D_a \psi \ e^J\wedge e^K\wedge e^L. \ \ 
\ea
Thus $n_e$ is essentially a discretised \emph{controvariant} tetrad vector field $e^a_I$. In a clean geometrical language this is actually a three-form, and is naturally integrated on 3d hyper-surfaces, whose 3d volume it determines. See  \cite{Wieland2014} and \cite{Cortes2014}.

\section{Lorentzian spectral triples and Dirac operator space}

The fermion space $\cal H$ defined above, the Dirac operator \eqref{D} and the commutative algebra $\cal A$ of complex  functions on $\cal V$ form a triple $C=({\cal A},{\cal H},D)$ which describes a discrete Lorentzian geometry.  This is easily generalised from chiral Weyl fermions to Dirac fermions, doubling fermion space and using Dirac's gamma matrices.  

The main difference with the standard definition of spectral triple is that $\cal H$ does not have a canonical Hilbert structure.  Rather, it has, so to say, a Hilbert structure for each choice of Lorentzian reference frame at each vertex.  From the point of view of physics the reason for this is transparent.  The standard notion of state space in quantum mechanics is associated to a spacelike surface in spacetime; the scalar product is not invariant under change of this surface. 

From the mathematical perspective, it is only when a particular $SU(2)$ subgroup of $SL(2,\C)$ is chosen that a scalar product in $\C^2$ is determined.  Not fixing a canonical scalar product for the fermion field may be needed for a proper physical definition of Lorentzian spectral geometry. 

The (generalised) spectral triple defined here is finite \cite{Krajewski1998}. It is also commutative, because it defines the (generalised) metric  structure of a single spinfoam.  The noncommutative aspects of quantum gravity, namely its quantum geometry, appear when summing over spinfoams, which is to say integrating over metrics. 

A particularly interesting aspect of the construction given is that it leads to a neat definition for the space of the Dirac operators.  A Dirac operator $D$ on a two--complex is determined by assigning a timelike Minkowski vector to each edge.  Therefore the space of Dirac operators is 
\be
{\cal D}=\otimes_e\ M^{+},
\ee
the product of $E$ times the cone $M^+ \subset R^{3,1}$ formed by the vectors $n$ with $|n|=n_{00}-|\vec n|^2>0$.  

I leave two questions open.  First, whether the ``spectral action" of these operators
\be
   S={\rm Tr}[\chi(D)]
\ee
where $\chi$ is the characteristic function of the interval $[-1,1]$ approximates a discretisation the Einstein Hilbert action. Given the remarkable analogous result in the continuum \cite{Chamseddine:1991qh}, this is not impossible and could provide an interesting step towards the construction of the spectral formulation of Lorentzian general relativity.  Second, whether an integration over the finite dimensional space $\cal D$ can be interpreted as  an integration over geometries.   If so, this would directly tie the spinfoam quantum dynamics with the natural quantisation of Connes' spectral geometry. 

\vspace{.5cm}
\centerline{---}
\vspace{.5cm}

I thank Ali Chamseddine for a conversation and especially Paolo Bertozzini for extensive help.

 \end{document}